\documentclass[prd,nofootinbib,superscriptaddress,tightenlines,notitlepage,floatfix]{revtex4-1}
\usepackage{bm}
\usepackage{tensor}
\usepackage{graphicx}
\usepackage{amssymb}
\usepackage{amsmath}
\usepackage{bm}
\usepackage{xcolor}
\usepackage{hyperref}
\usepackage{enumitem}
\usepackage{soul}  

\def\al{\alpha}
\def\be{\beta}
\def\ga{\gamma}
\def\de{\delta}
\def\ep{\epsilon}

\def\th{\theta}

\def\rh{\rho}

\def\ph{\phi}

\def\Ga{\Gamma}
\def\De{\Delta}

\def\Ph{\Phi}

\def\Om{\Omega}

\def\lsim{\mathrel{\rlap{\lower4pt\hbox{\hskip1pt$\sim$}}
    \raise1pt\hbox{$<$}}}
\def\gsim{\mathrel{\rlap{\lower4pt\hbox{\hskip1pt$\sim$}}
    \raise1pt\hbox{$>$}}}
\def\sqr#1#2{{\vcenter{\vbox{\hrule height.#2pt
         \hbox{\vrule width.#2pt height#1pt \kern#1pt
         \vrule width.#2pt}
         \hrule height.#2pt}}}}

\def\prt{\partial}
\def\lrpartial{\raise 1pt\hbox{$\stackrel\leftrightarrow\partial$}}

\def\part2{\partial_\alpha \partial^\alpha}

\def\xx'{|\vec x -\vec x'|}

\def\b2{b^\al b_\al}

\newcommand{\beq}{\begin{equation}}
\newcommand{\eeq}{\end{equation}}
\newcommand{\bea}{\begin{eqnarray}}
\newcommand{\eea}{\end{eqnarray}}
\newcommand{\bit}{\begin{itemize}}
\newcommand{\eit}{\end{itemize}}
\newcommand{\rf}[1]{(\ref{#1})}

\newcommand{\DOA}{\affiliation{Department of Astronomy, School of Physics,
Peking University, Beijing 100871, China}}
\newcommand{\KIAA}{\affiliation{Kavli Institute for Astronomy and
Astrophysics, Peking University, Beijing 100871, China}}
\newcommand{\NAOC}{\affiliation{National Astronomical Observatories,
Chinese Academy of Sciences, Beijing 100012, China}}

\begin{document}
\title{Signature of Lorentz Violation in Continuous Gravitational-Wave
Spectra of Ellipsoidal Neutron Stars}\thanks{Invited article to special
issue ``Lorentz Violation in Astroparticles and Gravitational Waves'' in
journal {\it Galaxies}.}
\author{Rui Xu}\email[Corresponding author: ]{xuru@pku.edu.cn}\KIAA
\author{Yong Gao}\DOA\KIAA
\author{Lijing Shao}\email[Corresponding author: ]{lshao@pku.edu.cn}\KIAA\NAOC
\date{\today}

\begin{abstract}
We study effects of Lorentz-invariance violation on the rotation of neutron
stars (NSs) in the minimal gravitational Standard-Model Extension
framework, and calculate the quadrupole radiation generated by them. Aiming
at testing Lorentz invariance with observations of continuous gravitational
waves (GWs) from rotating NSs in the future, we compare the GW spectra of a
rotating ellipsoidal NS under Lorentz-violating gravity with those of a
Lorentz-invariant one. The former are found to possess frequency components
higher than the second harmonic, which does not happen for the latter,
indicating those higher frequency components to be potential signatures of
Lorentz violation in continuous GW spectra of rotating NSs.
\end{abstract}
\maketitle

\section{Introduction}

The observation of gravitational waves (GWs) from the compact binary system
GW170817 initiates the era of multimessenger astronomy \cite{GBM:2017lvd,
Monitor:2017mdv}. Gravitational theories, including the renowned general
relativity (GR), are exposed to unprecedented tests utilizing GW
signals \cite{Corda:2009re}. Lorentz invariance, incorporated locally in GR as well as many
other alternative gravitational theories, is certainly one of the
fundamental principles subjected to these tests~\cite{Mirshekari:2011yq,
Kostelecky:2016kfm, Shao:2020shv, Liu:2020slm, Xu:2019gua}. By employing
the Standard-Model Extension (SME) framework \cite{Colladay:1996iz,
Colladay:1998fq, Kostelecky:2002hh, Kostelecky:2003fs, Bailey:2006fd},
which is widely used to investigate consequences from possible violations
of Lorentz invariance in terrestrial experiments and astrophysical
observations \cite{Kostelecky:2008ts}, stringent bounds have been set for
the coefficients for Lorentz violation in the gravitational sector of the
SME framework after analyzing the observed GW data
\cite{Kostelecky:2016kfm, Monitor:2017mdv, Shao:2020shv, Liu:2020slm}.

Besides the coalescence of compact binary systems, another type of GW
sources are deformed rotating neutron stars (NSs). Especially, when the
angular velocity of a deformed NS is misaligned with its angular momentum,
the star precesses about the direction of the angular momentum, radiating
out GWs continuously \cite{Zimmermann:1979ip, Zimmermann:1980ba}. The
search for such continuous GW signals is ongoing \cite{Pisarski:2019vxw,
Covas:2020nwy, Dergachev:2020fli, Papa:2020vfz, Steltner:2020hfd,
Zhang:2020rph}. Once detected, the continuous GW signals will tell us a
substantial piece of information on NS structure and deformability.
Furthermore, they will bring new tests for the laws of physics, among which
lies Lorentz invariance as one of the fundamental principles (see
e.g.~Ref.~\cite{Xu:2019gua}).

To test Lorentz invariance, an investigation of the scenario where it is
violated is necessary. The effects of Lorentz violation on rotating
spheroidal stars are studied in detail in Ref.~\cite{Xu:2020zxs} under the
minimal gravitational SME framework. The modification to the free
precession of a deformed star is depicted by the name {\it{twofold
precession}}, as briefly speaking, Lorentz violation causes the angular
momentum to precess about a fixed direction while at the same time the star
still precesses about the instantaneous direction of the angular momentum.
The correction in the quadrupole radiation due to the modification of the
rotation of the star is calculated in Ref.~\cite{Xu:2020zxs}, and it is
found that the quadrupole radiation from a spheroidal star affected by
Lorentz violation has frequency components higher than twice of the
fundamental one.

In this work, we are going to extend the numerical results in
Ref.~\cite{Xu:2020zxs} to ellipsoidal NSs. The characteristic higher
harmonics due to Lorentz violation remain in the GW spectra as we expect.
But more importantly, our numerical calculation for the quadrupole
radiation from an ellipsoidal NS in the absence of Lorentz violation
indicates that though the nonaxisymmetry of the star modulates the first
and the second harmonics in the GW spectra as discussed in
Refs.~\cite{Zimmermann:1980ba, VanDenBroeck:2004wj, Gao:2020zcd}, it does
not generate harmonics higher than the second for freely precessing NSs.
Therefore, harmonics higher than the second are indeed possible signatures
for Lorentz violation in the GW spectra of rotating solitary NSs.

We organize the paper as follows. In Sec.~\ref{sec2}, we present the
analytical equations to construct the quadrupole radiation from a rotating
ellipsoid under Lorentz-violating gravity. Then in Sec.~\ref{sec3},
numerical solutions to the rotation equations for ellipsoids with uniform
density are obtained and used to construct examples of the quadrupole
radiation. Subsequently, Fourier transformations are performed to extract the
frequency components of the quadrupole GWs, and we will see that while the
GW from an ellipsoid under the twofold precession contains harmonics higher
than the second, the GW from an ellipsoid under free precession only has
frequencies around the first and the second harmonics. In the end,
conclusions are summarized in Sec.~\ref{sec4}. For simplicity in writing
equations, we use the geometrized unit system where $G = c = 1$. However,
standard units do appear when numerical estimations are desired for
realistic NSs.

\section{Theoretical basics}
\label{sec2}

To proceed with the calculation, we neglect relativistic corrections to NS
structure and motion, and solve its motion from the rotation equations for
rigid bodies.\footnote{Relativistic corrections are reasonably
characterized by the compactness of the body, which is about $ 0.1 $ for a
NS. } Assuming that in the body frame $x$-$y$-$z$, the surface of the star
is described by
\bea
\frac{x^2}{a_x^2} + \frac{y^2}{a_y^2} + \frac{z^2}{a_z^2} = 1 ,
\eea
with semi-axes $a_x, \, a_y$ and $a_z$, then the Lagrangian for the
rotation of the star can be written as
\bea
L = \frac{1}{2} \left( I^{xx} (\Om^x)^2 + I^{yy} (\Om^y)^2 + I^{zz} (\Om^z)^2\right) - \de U ,
\label{lagr}
\eea
where $I^{xx}, \, I^{yy}$ and $I^{zz}$ are the eigenvalues of the moment of
inertia tensor along the principal axes, and $\Om^x, \, \Om^y $ and $\Om^z$
are the components of the angular velocity of the star in the $x$-$y$-$z$
frame. The orientation-dependent self-energy $\de U$ is calculated from the
anisotropic correction $\de \Phi$ to the Newtonian potential $\Phi$ in the
minimal gravitational SME, namely \cite{Bailey:2006fd} 
\bea
\de \Phi &=& -\frac{1}{2} \bar s^{ij} \int \frac{ (x^i-x^{\prime\,i}) (x^j - x^{\prime\,j}) }{|{\boldsymbol{x}} - {\boldsymbol{x}'}|^3} \rh({\boldsymbol{x}'}) \mathrm{d}^3\boldsymbol{x}',
\nonumber \\
\de U &=& -\frac{1}{4} \bar s^{ij} \int \frac{ \left( x^i - x^{\prime\,i} \right) \left( x^j - x^{\prime\,j} \right) }{|\boldsymbol{x} - \boldsymbol{x}'|^3} \rh(\boldsymbol{x}) \rh(\boldsymbol{x}') \mathrm{d}^3 \boldsymbol{x} \, \mathrm{d}^3 \boldsymbol{x}' ,
\eea
where $\bar s^{ij}$, with $i, j = x, y, z$, are the coefficients for
Lorentz violation in the body frame~\cite{Kostelecky:2003fs,Bailey:2006fd},
and $\rh$ is the density of the star.

\begin{figure*}
    \centering
    \includegraphics[width=6cm]{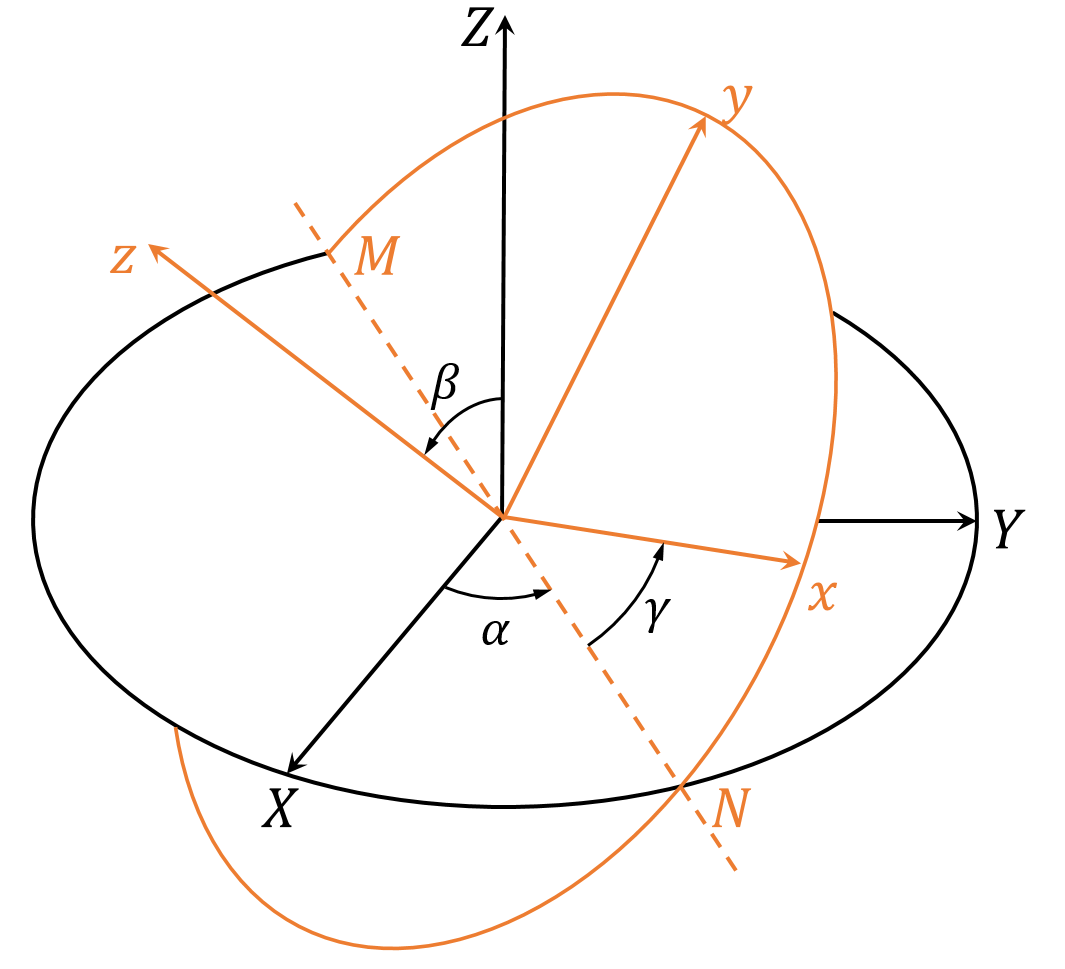}
    \caption{Euler angles transforming the $X$-$Y$-$Z$ inertial frame to
    the $x$-$y$-$z$ body frame. First, rotate the $X$-$Y$-$Z$ frame about
    the $Z$ axis with angle $\al$ so that the $X$-axis aligns with the
    intersection line $MN$. Then, rotate the just obtained $X$-$Y$-$Z$
    frame about the line $MN$ with angle $\be$ so that the $Z$-axis aligns
    with the $z$-axis. Last, rotate the new $X$-$Y$-$Z$ frame about the
    $z$-axis with angle $\ga$ so that it overlaps with the $x$-$y$-$z$
    frame. }
    \label{euler}
\end{figure*}

In the SME framework, the coefficients for Lorentz violation are assumed to
be constant in inertial frames. Therefore, as the star rotates, the
coefficients $\bar s^{ij}$ depend on the orientation of the star according
to
\bea
\bar s^{ij} = R^{iI} R^{jJ} \bar s^{IJ} ,
\eea
where $R^{iI}$ represents the rotation matrix transforming an inertial
frame $X$-$Y$-$Z$ to the body frame $x$-$y$-$z$. The capital indices run
over $X, \, Y$ and $Z$, and $\bar s^{IJ}$ are constant coefficients for
Lorentz violation. The orientation dependence of $\bar s^{ij}$, originated
from the rotation matrix, can be easily described by the Euler angles
$(\al, \be, \ga)$ in Fig.~\ref{euler}, as the rotation matrix in terms of
the Euler angles is
\begin{equation}
\begin{array}{@{}c@{\hspace{1ex}}c@{}}
 & \text{ $\xrightarrow{\hspace*{5cm}I\hspace*{5cm}}$ } \\[1ex]
R^{iI} =
\rotatebox[x=0cm,y=-0.6cm]{90}{\text{$\xleftarrow{\hspace*{0.5cm}i\hspace*{0.5cm}}$}} &
\begin{pmatrix}
 \cos\al\cos\ga-\sin\al\cos\be\sin\ga & \sin\al\cos\ga+\cos\al\cos\be\sin\ga & \sin\be\sin\ga \\
 -\cos\al\sin\ga-\sin\al\cos\be\cos\ga & -\sin\al\sin\ga+\cos\al\cos\be\cos\ga & \sin\be\cos\ga \\
 \sin\al\sin\be & - \cos\al\sin\be & \cos\be 
 \end{pmatrix} .
\end{array} 
\end{equation}
Together with the relations between the velocity components and the Euler
angles \cite{landau1960course},
\bea
\Om^x &=& \dot \al \sin\be \sin\ga + \dot \be \cos\ga ,
\nonumber \\
\Om^y &=& \dot \al \sin\be \cos\ga - \dot \be \sin\ga ,
\nonumber \\
\Om^z &=& \dot \al \cos\be + \dot \ga,
\label{angvel}
\eea
where dots denote time derivatives, the Euler-Lagrange equations for the
Euler angles can be obtained from the Lagrangian \rf{lagr}. Given the shape
and density of the star, the moment of inertia tensor and the integrals in
$\de U$ can be calculated, and then the Euler angles can be solved to
describe the rotation of the star.

Once the rotation of the star is known, its gravitational quadrupole
radiation can be calculated via the metric perturbation
\bea
h^{IJ} = - \frac{2}{r} \ddot I^{IJ} ,
\eea
where $r$ is the distance from the distant star to the observer, and the double dots denote the second time derivative. In Ref.~\cite{Xu:2020zxs}, it is shown that $\ddot I^{IJ}$ can be written as
\bea
\ddot I^{IJ} = R^{iI} R^{jJ} A^{ij} ,
\eea
with the body-frame quantities
$A^{ij}$ being
\bea
A^{xx} &=& 2\left( \De_2 \left(\Om^y\right)^2 - \De_3 \left(\Om^z\right)^2 \right) ,
\nonumber \\
A^{yy} &=& 2\left( \De_3 \left(\Om^z\right)^2 - \De_1 \left(\Om^x\right)^2 \right) ,
\nonumber \\
A^{zz} &=& 2\left( \De_1 \left(\Om^x\right)^2 - \De_2 \left(\Om^y\right)^2 \right) ,
\nonumber \\
A^{xy} &=& \left( \frac{ \left(\De_3 \right)^2}{I^{zz}} + \De_1 - \De_2 \right) \Om^x \Om^y + \frac{\De_3}{I^{zz}} \Ga^z,
\nonumber \\
A^{xz} &=& \left( \frac{ \left(\De_2 \right)^2}{I^{yy}} + \De_3 - \De_1 \right) \Om^x \Om^z + \frac{\De_2}{I^{yy}} \Ga^y ,
\nonumber \\
A^{yz} &=& \left( \frac{ \left(\De_1 \right)^2}{I^{xx}} + \De_2 - \De_3 \right) \Om^y \Om^z + \frac{\De_1}{I^{xx}} \Ga^x ,
\label{gwa}
\eea
for any rigid body subjected to arbitrary rotations. The quantities $\De_1, \, \De_2$ and $\De_3$ are defined as
\bea
\De_1 = I^{yy}-I^{zz}, \quad \De_2 = I^{zz}-I^{xx}, \quad \De_3 = I^{xx}-I^{yy},
\eea
and the components of the torque, $\Ga^x, \, \Ga^y$ and $\Ga^z$, are calculated from the orientation-dependent self-energy $\de U$ via
\bea
\Ga^x &=& -\frac{\sin\ga}{\sin\be} \, \prt_\al \de U - \cos\ga \, \prt_\be \de U + \cot\be\sin\ga \, \prt_\ga \de U, 
\nonumber \\
\Ga^y &=& -\frac{\cos\ga}{\sin\be} \, \prt_\al \de U + \sin\ga \, \prt_\be \de U + \cot\be\cos\ga \, \prt_\ga \de U, 
\nonumber \\
\Ga^z &=& - \prt_\ga \de U .
\eea

Finally, the two physical degrees of freedom in the GW can be extracted
from $h^{IJ}$ by defining the plus and the cross modes for an observer
whose colatitude and azimuth are $\th_o$ and $\ph_o$ in the $X$-$Y$-$Z$
frame \cite{Poisson:2014},
\bea
h_{+} &=& \frac{1}{2} \left( \hat {\th}_o^I \, \hat \th_o^J  - \hat {\ph}_o^I \, \hat {\ph}_o^J \right) h^{IJ} = - \frac{1}{r} \left( \hat {\th}_o^I \, \hat \th_o^J  - \hat {\ph}_o^I \, \hat {\ph}_o^J \right) \ddot I^{IJ}, 
\nonumber \\
h_{\times} &=& \frac{1}{2} \left( \hat {\th}_o^I \, \hat \ph_o^J  + \hat {\ph}_o^I \, \hat {\th}_o^J \right) h^{IJ} = - \frac{1}{r} \left( \hat {\th}_o^I \, \hat \ph_o^J  + \hat {\ph}_o^I \, \hat {\th}_o^J \right) \ddot I^{IJ},
\label{hpluscross}
\eea
where $\hat \th_o^I$ and $\hat \ph_o^I$ are the $XYZ$-components of the transverse unit vectors 
\bea
\hat {\boldsymbol{\th}}_o &=& \cos\th_o \cos\ph_o \, \hat{\boldsymbol{e}}_X + \cos\th_o \sin\ph_o \, \hat{\boldsymbol{e}}_Y - \sin\th_o \hat{\boldsymbol{e}}_Z,
\nonumber \\
\hat {\boldsymbol{\ph}}_o &=& -\sin\ph_o \, \hat{\boldsymbol{e}}_X + \cos\ph_o \, \hat{\boldsymbol{e}}_Y  .
\label{thphinertial}
\eea
Note that $\hat{\boldsymbol{e}}_X, \, \hat{\boldsymbol{e}}_Y$ and $\hat{\boldsymbol{e}}_Z$ are the unit vectors of the $X$-$Y$-$Z$ frame, while $\hat{\boldsymbol{e}}_x, \, \hat{\boldsymbol{e}}_y$ and $\hat{\boldsymbol{e}}_z$ will be used as the unit vectors of the $x$-$y$-$z$ frame.

\section{Numerical examples}
\label{sec3}

Now we can use the above equations to numerically calculate the GW spectra
of a rotating ellipsoidal NS affected by Lorentz violation. To simplify the
calculation of $\de U$, we assume the density of the star to be constant.
Extension to realistic nonuniform NSs is straightforward. Then the angular
parts of the integrals in $\de U$ can be carried out analytically.
Specifically speaking, define
\bea
U^{ij} := \frac{1}{2} \int \frac{ \left( x^i - x^{\prime\,i} \right) \left( x^j - x^{\prime\,j} \right) }{|\boldsymbol{x} - \boldsymbol{x}'|^3} \rh(\boldsymbol{x}) \rh(\boldsymbol{x}') \mathrm{d}^3\boldsymbol{x} \, \mathrm{d}^3\boldsymbol{x}' ,
\eea
then they are related to the Newtonian potential
\bea
\Ph = - \int \frac{ \rh(\boldsymbol{x}') }{|\boldsymbol{x} - \boldsymbol{x}'|} \mathrm{d}^3\boldsymbol{x}' ,
\eea
via
\bea
U^{ij} &=& \int \rh (\boldsymbol{x}) x^i \prt_j \Phi \mathrm{d}^3 \boldsymbol{x} .
\eea
The Newtonian potential of a uniform ellipsoid is known to be \cite{1962ApJ...136.1037C, Poisson:2014}
\bea
\Ph = -\pi \rh \left( A_0 - A_x x^2 - A_y y^2 - A_z z^2 \right), 
\eea
where
\bea
A_0 = a_x a_y a_z \int_0^\infty \frac{\mathrm{d}u}{ \sqrt{(a_x^2+u)(a_y^2+u)(a_z^2 + u)} } , \quad 
A_i = a_x a_y a_z \int_0^\infty \frac{\mathrm{d}u}{ (a_i^2 + u) \sqrt{(a_x^2+u)(a_y^2+u)(a_z^2 + u)} } ,
\eea
with $i = x, y, z$. Consequently, the nonvanishing $U^{ij}$ are found to be
\bea
U^{xx} = \frac{8\pi^2}{15} \rh^2 A_x a_x^3 a_y a_z, \quad U^{yy} = \frac{8\pi^2}{15} \rh^2 A_y a_x a_y^3 a_z, \quad U^{zz} = \frac{8\pi^2}{15} \rh^2 A_z a_x a_y a_z^3. 
\eea

For NSs, the density varies from the center to the surface. For our
purpose, we will take a uniform density of $10^{15} \, {\rm {g/cm}^3}$ in
numerical calculations. As for the semi-axes, because NSs are compact objects having tiny deformations if any, we can only say that they are all
about $10 \, {\rm km}$, roughly the radius of a spherical NS predicted by GR. The often
used parameters to characterize NS deformation are the oblateness $\ep$ and
the nonaxisymmetry $\de$. They are defined as
\bea
\ep = \frac{ I^{zz} - I^{xx} }{ I^{xx} }, \quad \de = \frac{ I^{yy} - I^{xx}  }{ I^{zz} - I^{xx} },
\eea
with an assumption that $I^{zz}$ is the largest eigenvalue of the moment of
inertia tensor. NS models have suggested that $\ep$ is less than $10^{-7}$
\cite{Owen:2005fn}, while the magnitude of $\de$ is hardly known. For
demonstration, we take 0.1 for both $\ep$ and $\de$ in the following
numerical examples. In addition, we use $10 \, {\rm km}$ for $a_z$, and
then the values of $a_x$ and $a_y$ are determined by noticing
\bea
I^{xx} = \frac{4\pi}{15} \rh a_xa_ya_z(a_y^2+a_z^2) , \quad I^{yy} = \frac{4\pi}{15} \rh a_xa_ya_z(a_x^2+a_z^2) , \quad I^{zz} = \frac{4\pi}{15} \rh a_xa_ya_z(a_x^2+a_y^2) ,
\eea 
for uniform ellipsoids.

\begin{figure*}
    \centering
    \includegraphics[width=16cm]{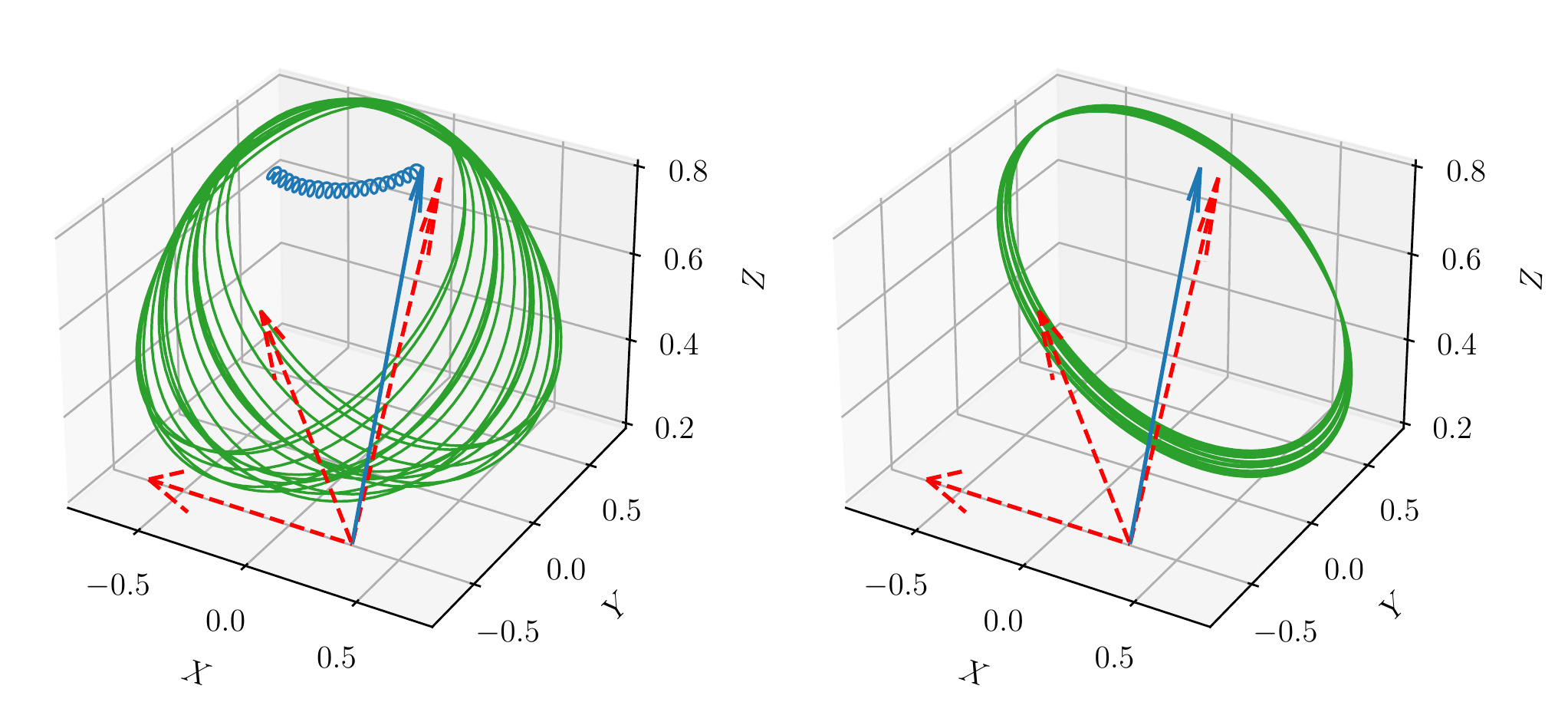}
    \caption{Illustrations for an example of the Lorentz-violating twofold
    precession ({\it left}) and an example of the Lorentz-invariant free
    precession ({\it right}). The green trajectories trace the tail of the
    body-frame unit vector $\hat {\boldsymbol{e}}_z$ in the inertial frame,
    while the blue trajectory in the left plot traces the tail of the
    angular momentum unit vector in the inertial frame. The red arrows are
    the body-frame unit vectors $\hat {\boldsymbol{e}}_x , \, \hat
    {\boldsymbol{e}}_y$ and $\hat {\boldsymbol{e}}_z$ at $t=0$, while the
    blue arrows indicate the angular momentum unit vector at $t=0$. The angular
    momentum is conserved in free precessions so the blue arrow in the
    right plot remains unchanged with time. The initial values for both
    solutions are $\al|_{t=0}=0$, $\be|_{t=0} \approx 0.762$, $\ga|_{t=0}
    =\pi/2$, and $\dot \al|_{t=0} \approx 1.26$, $\dot \be|_{t=0} = 0.5$,
    $\dot \ga|_{t=0} \approx 0.0449$. Time and time derivatives are
    dimensionless under the time unit $t_c$ given by Eq.~\rf{tu1}. }
    \label{fig1}
\end{figure*}

\begin{figure*}
    \centering
    \includegraphics[width=18cm]{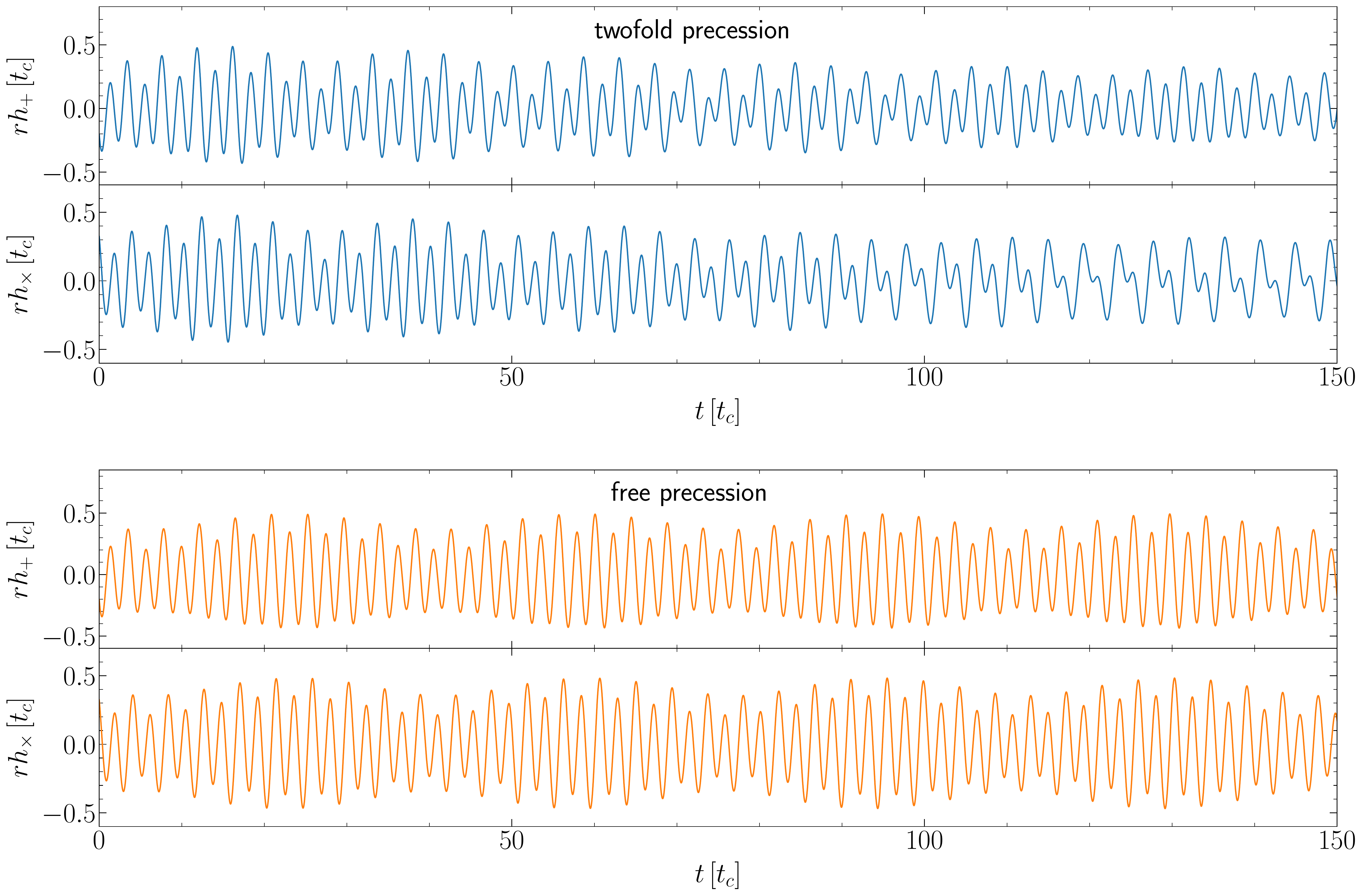}
    \caption{GWs from a rigid body undergoing the rotations in Fig.~\ref{fig1}. The observer receiving the waves has colatitude $\th_o = 0.8 \, {\rm rad}$ and azimuth $\ph_o = 0$ in the $X$-$Y$-$Z$ frame. The geometrized unit of time and distance is the time unit $t_c$ given by Eq.~\rf{tu1}.  }
    \label{fig2}
\end{figure*}

\begin{figure*}
    \centering
    \includegraphics[width=16cm]{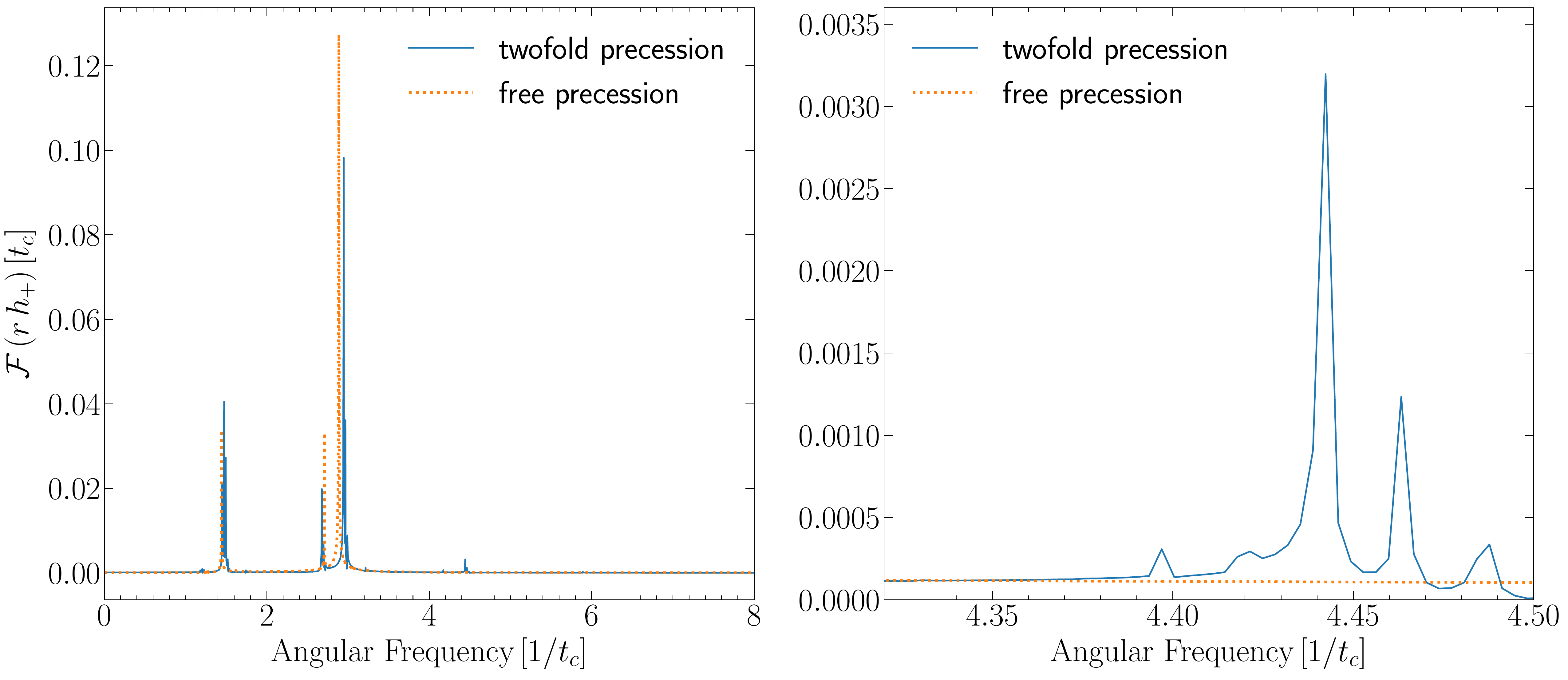}
    \caption{Fourier transformations of the $r h_+$ waves in Fig.~\ref{fig2}. The two noticeable peaks at about 1.4 and 2.9 in the left plot are the first and the second harmonics for both twofold precession and free precession. The modulation due to nonaxisymmetry is clearly represented by the adjacent peak at about 2.7 close to the second harmonic for both kinds of motion. However, the barely visible tiny peaks, reflecting modulations due to Lorentz violation, only exist for twofold precession. The right plot, which zooms in on the tiny peak between 4 and 5, demonstrates the point. Note that in the plots the geometrized unit of Fourier amplitude is $t_c$, while the geometrized unit of angular frequency is $1/t_c$. }
    \label{fig3}
\end{figure*}

Then to compute $\de U$ as a function of the Euler angles, we take numerical values $\bar s^{XX} = 0.02, \, \bar s^{YY}=0.01, \, \bar s^{ZZ} = -0.04$ and $\bar s^{XY} = \bar s^{XZ} = \bar s^{YZ} = 0$ for the coefficients for Lorentz violation in the inertial frame. This means that the axes of the inertial frame are the principal axes of the $\bar s^{ij}$ tensor. Note that this is a theoretical inertial frame fixed by the coefficients for Lorentz violation. It generally does not coincide with the widely used experimental inertial frame, namely the Sun-centered celestial-equatorial frame defined in Ref.~\cite{Kostelecky:2002hh}. 

All the parameters in the Lagrangian \rf{lagr} have been set now. Numerical solutions for the Euler angles can be obtained once initial values are given. For numerical calculations, a dimensionless parametrization for the angular velocities is helpful. This can be achieved by employing a time unit. To be consistent with the choice in Ref.~\cite{Xu:2020zxs}, it is taken to be
\bea
t_c := \sqrt{ \frac{ 2I^{xx} }{ U^{yy} - U^{zz} } } .
\eea 
For a uniform ellipsoid, keeping only the leading contribution from $\ep$ and $\de$, it is
\bea
t_c = \sqrt{ \frac{a_y^2 + a_z^2}{\pi \rh \left(A_y a_y^2 - A_z a_z^2 \right) } } \approx \sqrt { \frac{15}{4\pi \rh \ep (1-\de) } } \sim 10^{-3} \, {\rm s} ,
\label{tu1}
\eea 
where the magnitude estimation is made for $\rh = 10^{15} \, {\rm g/cm}^3$ and $\ep = \de = 0.1$. Therefore, a dimensionless angular velocity at order unity in our numerical results corresponds to about $1000 \, {\rm rad/s}$.

Figure \ref{fig1} shows the trajectories of the tail of the unit vector $\hat{ \bm{e}}_z$ in the inertial frame to intuitively illustrate the rotations of the star for a certain set of initial values. Our examples consist of two solutions: the plot on the left shows a twofold precession with $\bar s^{IJ}$ taking the above said values, and the plot on the right shows a free precession without Lorentz violation for comparison. The distinction is also reflected by the trajectories of the tail of the angular momentum unit vector: in the left plot, there is a nontrivial trajectory for the angular momentum unit vector, while in the right plot, the angular momentum unit vector does not change with time.    

With the two solutions, we calculate the GWs according to Eq.~\rf{hpluscross} for an observer at $\th_o = 0.8 \, {\rm rad}$ and $\ph_o = 0$. The results are presented in Fig.~\ref{fig2}. Their Fourier transformations are shown in Fig.~\ref{fig3}; only the plus mode is shown as the cross mode has very much the same spectra. The spectra of the free precession shows a fundamental angular frequency at about $1.4/t_c$, and peaks around the second harmonic at about $2.9/t_c$. We know that if the star is axisymmetric, free precessions generate GWs having exactly two frequencies, with one being twice of the other. The nonaxisymmetry here modulates both the fundamental frequency and the second harmonic. This has been discussed in Refs.~\cite{Zimmermann:1980ba, VanDenBroeck:2004wj, Gao:2020zcd}. What we are showing in the left plot of Fig.~\ref{fig3} tells us that the twofold precession, namely the rotation of an otherwise freely precessing NS under Lorentz-violating gravity, generates similar GW frequency components. However, more interestingly, in the enlarged plot on the right, we clearly see the distinction that while the twofold precession generates frequency components around the third harmonic, the free precession has no component of the third harmonic at all. Higher frequency components exist in the spectra of the Lorentz-violating twofold precession, but they can easily be missed as they are too small.

\section{Conclusion}
\label{sec4}

We have presented the analytical formulae to calculate the rotation of NSs
under Lorentz-violating gravity in the minimal gravitational SME framework,
and to construct the quadrupole GWs emitted from these NSs. Numerical
examples are plotted to demonstrate our conclusion that while freely
precessing NSs in the Lorentz-invariant gravity do not emit quadrupole GWs
at frequencies higher than the second harmonic, NSs undergoing the twofold
precession due to Lorentz violation do. Therefore, harmonics higher than
the second in the spectra of continuous GWs are appealing signatures of
Lorentz violation. Once continuous GWs from rotating NSs are detected, a
potential test of Lorentz invariance can be performed by examining
harmonics higher than the second in the spectra. However, we do notice a
possible difficulty in this test: there might be conventional torques, like
the electromagnetic spin-down torque \cite{1970ApJ...160L..11G,
Jones:2001yg, Zanazzi:2015ida, Gao:2020dwy}, acting on the NS to cause
similar twofold precession motions and to generate higher harmonics in the
GW spectra. Although the questions whether the twofold precession caused by Lorentz
violation can be distinguished from rotations of NSs under the
electromagnetic spin-down torque and whether the GW spectra of the latter
have frequency components higher than the second harmonic lie beyond the
scope of this work, they are certainly worth to be investigated further. Furthermore, a statistical study of continuous GWs from an
ensemble of NSs might have the potential to distinguish between the two
scenarios, as the Lorentz violation is universal for all NSs while the
astrophysical torques are different for different systems.

\begin{acknowledgements}
We are grateful to Marco Schreck for the invitation to submit an article to this special issue.
This work was supported by the National Natural Science Foundation of China
(11975027, 11991053, 11721303), the National SKA Program of China (2020SKA0120300), the Young Elite Scientists Sponsorship
Program by the China Association for Science and Technology (2018QNRC001),
the Max Planck Partner Group Program funded by the Max Planck Society, and
the High-Performance Computing Platform of Peking University.
It was partially supported by the Strategic Priority Research Program of
the Chinese Academy of Sciences through the Grant No. XDB23010200.
R.X. is supported by the Boya Postdoctoral Fellowship at Peking University.
\end{acknowledgements}

\bibliography{refs}

\end{document}